\documentstyle[preprint,aps,tighten]{revtex}

\begin{document}
\draft
\title{Multi-electron SEFs for nuclear reactions involved in advanced stages of
stellar evolution}
\author{Theodore E. Liolios $^{1,2,3}$ \footnote{theoliol@physics.auth.gr}}
\address{$^1$European Center for Theoretical Studies in Nuclear Physics and Related Areas\\
Villa Tambosi, I-38050 Villazzano (Trento), Italy\\
\footnote{Correspondence address}$^2$University of Thessaloniki, Department of Theoretical Physics\\
Thessaloniki 54006, Greece\\
$^3$ Hellenic War College, BST 903, Greece\\
}

\maketitle

\begin{abstract}
Multi-electron screening effects encountered in laboratory astrophysical
reactions are investigated by considering the reactants Thomas-Fermi atoms.
By means of that model , previous studies are extended to derive the
corresponding screening enhancement factor (SEF), so that it takes into
account ionization, thermal, exchange and relativistic effects. The present
study, by imposing a very satisfactory constraint on the possible values of
the screening energies and the respective SEFs, corrects the current (and
the future) experimental values of the astrophysical factors associated with
nuclear reactions involved in advanced stages of stellar evolution.
\end{abstract}


\section{Introduction}

The astrophysical factors for the nuclear reactions of the CNO bi-cycle
(solar center, upper main-sequence stars etc.) and of other advanced nuclear
burning stages, such as those ignited in classical novae, have been obtained
by performing measurements well above the Gamow peak, (\cite{rolfsbook} and
references therein) and extrapolating to lower energies thus committing a
significant error in certain cases. After the accomplishments of the LUNA
collaboration\cite{lunasecond} in probing deep into the Gamow-peak region of
the most important break-up reaction $He^{3}\left( He^{3},2p\right) He^{4}$
the nuclear astrophysics community hopes that similar low-energy experiments
will follow for astrophysical nuclear reactions involving heavier nuclei.
However, the screening effect, which has been studied extensively (\cite
{lioliosluna,shoppamolecular} and references therein) regarding nuclear
reaction of the proton-proton chain, is particularly accentuated when
heavier ions are considered due to their multi-electron nature. It is the
purpose of this paper to investigate the enhancement of the astrophysical
factor that appears (or will appear) when astrophysical energies are
attained in nuclear reaction experiments.

In a recent paper\cite{lioliosluna} the Thomas-Fermi (TF) model was used in
order to derive a lower (sudden) limit (SL) and an upper (adiabatic) limit
(AL), thus imposing a constraint on the possible values of the screening
energies. In the present work we will improve that study by tightening that
constraint while studying, at the same time, ionization, relativistic,
thermal and exchange effects.

The TF model is one of the most thorougly studied models of atomic physics
(for a detailed study see Ref. \cite{marchbook}). In brief, if $n\left(
r\right) $ is the electron number density around the screened nucleus then
the self-consistent potential $\Phi \left( r\right) ,$ which results from
the combined field (nucleus plus electrons), is obtained by solving the
Poisson equation $\nabla ^{2}\Phi \left( r\right) =-4\pi \rho \left(
r\right) $ where $\rho \left( r\right) =-en\left( r\right) $ is the charge
density of the electron cloud. Using quantum statistical arguments\cite
{marchbook} we can show that for the TF atom at each point $r$ the fermi
energy $p_{f}$ must be

\begin{equation}
\frac{p_{f}^{2}}{2m_{e}}=e\left[ \Phi \left( r\right) -\Phi \left(
r_{0}\right) \right]
\end{equation}
where $r_{0}$ is the ionic radius which gives the boundary conditions

\begin{equation}
\Phi \left( r_{0}\right) =\frac{Zeq}{r_{0}},\left( \frac{d\Phi }{dr}\right)
_{r_{0}}=-\frac{\Phi \left( r_{0}\right) }{r_{0}}
\end{equation}
and the degree of ionization is a function of the number of electrons $%
\left( N\right) $ and protons $\left( Z\right) ,$ respectively, so that $%
q=1-N/Z$.

Following some well cited transformations\cite{marchbook} the
self-consistent potential can be written:

\begin{equation}
\Phi \left( r\right) =\Phi \left( r_{0}\right) +\frac{Ze}{r}\phi
\end{equation}
Setting $r=\alpha x,$ the universal function $\phi \left( x\right) $ is
given by the dimensionless TF equation:

\begin{equation}
\frac{d^{2}\phi \left( x\right) }{dx^{2}}=\frac{\phi ^{3/2}\left( x\right) }{%
\sqrt{x}}  \label{tf0}
\end{equation}
to be solved with the boundary conditions $\phi \left( 0\right) =1,$ $\phi
^{^{\prime }}\left( x_{0}\right) =-q/x_{0}.$ Note that the scaling parameter 
$\alpha $ has its usual value $\alpha =0.88534Z^{-1/3}a_{H},$ with $a_{H}$
representing the Bohr radius.

\section{Ionization, exchange and thermal effects}

We will exploit the advantages of the TF theory in order to derive the
effects of ionization, electron-electron (exchange) interactions and
temperature on multi-electron SEFs. The adiabatic limit (AL) SEF of Ref. 
\cite{lioliosluna} for neutral atoms seems to be well justified and it is
the largest available. However, the sudden limit (SL) SEF of the same work
needs improvement as it has been derived using Tietz's approximation\cite
{tietz} which is actually a rough model that can only approximate the
self-consistent potential of neutral atoms at average distances from the
nucleus, disregarding all relativistic, thermal and ionization effects. As
far as astrophysical reaction are concerned, screening becomes important
very close to the nucleus and therefore a more appropriate approximation is
needed, which could also account for ionization effects. To solve that
problem we resort to Baker's\cite{baker} small $x$ expansion where the
expansion coefficients are themselves functions of the initial slope $%
S\left( q\right) =\phi ^{^{\prime }}\left( 0\right) .$ The first few terms
are given by:

\begin{equation}
\phi _{B}\left( x\right) =1+S\left( q\right) x+\frac{4}{3}x^{3/2}+\frac{2}{5}%
S\left( q\right) x^{5/2}+\frac{1}{3}x^{3}+\frac{3}{70}S^{2}\left( q\right)
x^{7/2}+\frac{2}{15}S\left( q\right) x^{4}+...  \label{baker}
\end{equation}
There exists a unique value $S\left( 0\right) =-1.588$ which gives $%
x_{0}=\infty $ and $\phi \left( x_{0}\right) =0$ corresponding to the
neutral TF atom $\left( q=0\right) $. Every other value $S\left( q\right)
<S\left( 0\right) $ gives finite values for $x_{0}$ corresponding to
positive TF ions $\left( 0<q<1\right) .$ By selecting various initial slopes
we obtain relations of the form: $S=S\left( q\right) $ and $%
x_{0}=x_{0}\left( q\right) $ (see Figures 1 and 2)

Hence, the screened Coulomb potential around the target nucleus will now be
given by:

\begin{equation}
\Phi \left( r\right) =\frac{Z_{1}eq}{\alpha x_{0}\left( q\right) }+\frac{%
Z_{1}e}{r}\phi _{B}\left( \frac{r}{\alpha }\right)  \label{fion}
\end{equation}
where from now on $Z_{1}$ and $Z_{2}\,$represent the atomic numbers of the
target nucleus and the projectile, respectively

An alternative approach would be to adopt Moliere's approximation\cite
{moliere} for the solution of Eq. $\left( \ref{tf0}\right) $:

\begin{equation}
\phi _{M}\left( x\right) =\sum_{i=1}^{3}B_{i}\exp \left( -\beta _{i}x\right)
\label{mnr}
\end{equation}
with the fitting parameters 
\begin{equation}
B_{1}=0.1,\quad B_{2}=0.55,\quad B_{3}=0.35,\quad \beta _{1}=6.0,\quad \beta
_{2}=1.2,\quad \beta _{3}=0.3
\end{equation}

However, Moliere's function is less accurate than Baker's one since the
former underestimates the initial slope. For example, for neutral atoms $%
\phi _{M}^{^{\prime }}\left( 0\right) =-1.365,$ while $\phi _{B}^{^{\prime
}}\left( 0\right) =-1.588$

When positive ions are considered the total binding energy\cite{milne} $%
E_{TF}^{tot}=-20.93\,Z^{7/3}\,eV$ used in Ref.\cite{lioliosluna} for neutral
TF atoms must be modified to account for ionization. In such a case\cite{tal}

\begin{equation}
E_{TF}^{tot}\left( q\right) =F\left( q\right) Z^{7/3}  \label{totalq}
\end{equation}
where

\begin{equation}
F\left( q\right) =\frac{12}{7}\left( \frac{2}{9\pi ^{2}}\right) ^{1/3}\frac{%
e^{2}}{a_{H}}\left[ S\left( q\right) +\frac{q^{2}}{x_{0}\left( q\right) }%
\right]  \label{fq}
\end{equation}
We can now derive the screening enhancement factor for astrophysical nuclear
reactions. The energy dependent penetration factor $P\left( E\right) $
multiplied by the astrophysical factor $S\left( E\right) $ in the $s$-wave
cross section formula

\begin{equation}
\sigma \left( E\right) =\frac{S\left( E\right) }{E}P\left( E\right)
\label{cross}
\end{equation}
is given by the WKB method:

\begin{equation}
P^{SC}\left( E\right) =\exp \left[ -\frac{2\sqrt{2\mu }}{\hbar }%
\int_{r_{n}}^{r_{c}\left( E\right) }\sqrt{V_{sc}\left( r\right) -E}dr\right]
\label{pe}
\end{equation}
where $\left( SC\right) $ stands for screening and the classical turning
point $r_{c}$ is given by

\begin{equation}
V_{sc}\left( r_{c}\right) =E  \label{rc}
\end{equation}
while the nuclear radius $r_{n}\,$is a function of the mass numbers $A_{1,2}$
of the reactants, so that $r_{n}=1.4\left( A_{1}^{1/3}+A_{2}^{1/3}\right)
fm. $

At astrophysical energies in the laboratory the potential energy is found to
be shifted by a constant screening energy $U_{e}$ which is added to the
relative energy of the collision. For the no-screening (NOS) nucleus-nucleus
channel the calculation is easy leading to\cite{claytonbook}

\begin{equation}
P^{NOS}\left( E\right) =\exp \left[ -2\pi n\left( 1-\frac{4}{\pi }\left( 
\frac{r_{n}}{r_{c}}\right) ^{1/2}+\frac{2}{3\pi }\left( \frac{r_{n}}{r_{c}}%
\right) ^{3/2}+...\right) \right]
\end{equation}
where $n$ is the Sommerfeld parameter and $U_{e}=0$. The general SEF $%
f\left( E\right) \,$is defined as 
\begin{equation}
f\left( E\right) =\frac{P^{SC}\left( E\right) }{P^{NOS}\left( E\right) }
\label{fex}
\end{equation}
which for weak screening (WES) shifts $U_{e}<<E$ can be written\cite
{shoppaatomic}

\begin{equation}
f^{WES}\left( E\right) =\exp \left( \pi n\frac{U_{e}}{E}\right)  \label{fap}
\end{equation}
In fact all measurements in the laboratory should be corrected by means of
the above SEF so that the bare-nucleus astrophysical factor is not
overestimated. Following the treatment of Ref. \cite{lioliosluna} which was
used for neutral TF atoms, we can derive an approximate analytic formula for
the TF screening shifts $U_{TF}^{\left( SL,AL\right) }$and the corresponding
SEFs $f_{TF}^{\left( SL,AL\right) }\left( E\right) \,$in the SL and the AL
respectively, where the degree of ionization is also taken into account.
Combining Eqs. $\left( 16\right) ,\left( 12\right) ,$and $\left( 6\right) $
we obtain the following formulas:

Sudden Limit

\begin{equation}
U_{TF}^{SL}\left( q\right) =4\left( \frac{2}{9\pi ^{2}}\right)
^{1/3}Z_{1}^{4/3}Z_{2}\left[ S\left( q\right) +\frac{q}{x_{0}\left( q\right) 
}\right] \frac{e^{2}}{a_{H}}  \label{uslion}
\end{equation}

\begin{equation}
f_{TF}^{SL}\left( E\right) \simeq \exp \left[ -\frac{%
0.482Z_{1}^{7/3}Z_{2}^{2}A^{1/2}}{E_{\left( keV\right) }^{3/2}}\left[
S\left( q\right) +\frac{q}{x_{0}\left( q\right) }\right] \right]
\label{fslion}
\end{equation}

Adiabatic Limit

\begin{equation}
U_{TF}^{AL}=\left[ F\left( q_{12}\right) \left( Z_{1}+Z_{2}\right)
^{7/3}-F_{1}\left( q_{1}\right) Z_{1}^{7/3}-F_{2}\left( q_{2}\right)
Z_{2}^{7/3}\right]  \label{ualion}
\end{equation}

\begin{equation}
f_{TF}^{AL}\left( E\right) \simeq \exp \left[ -\frac{15.64Z_{1}Z_{2}\left[
F\left( q_{12}\right) \left( Z_{1}+Z_{2}\right) ^{7/3}-F_{1}\left(
q_{1}\right) Z_{1}^{7/3}-F_{2}\left( q_{2}\right) Z_{2}^{7/3}\right] A^{1/2}%
}{E_{\left( keV\right) }^{3/2}}\right]  \label{falion}
\end{equation}
where the subscripts $\left( 1,2,12\right) $ of $q$ $\,$in the adiabatic
limit formulas take into account the different degrees of ionization of the
reactants and the combined nuclear molecule.

The present sudden limit formulas are more accurate than the ones obtained
by means of Tietz's approximation, therefore improving the constraint on the
possible range of screening energies. For example the lower limit for
neutral atoms given by Eq. $\left( \ref{uslion}\right) \,$is $40\%$ larger
than the corresponding one of Ref.\cite{lioliosluna}$\,$, thus greatly
narrowing the gap between the sudden and the adiabatic limit SEFs.

Further simplifications are possible by noting that, for small ionization
numbers $q,$ the second additive term $q^{2}/x_{0}\left( q\text{ }\right) \,$%
in Eq. $\left( \ref{fq}\right) $ is much smaller than the initial slope $%
S\left( q\right) $ and can therefore be neglected$.$ If we further assume
proton induced nuclear fusion reactions where the ionization state before
and after the collision is the same then the $AL$ formulas can be written as

\begin{equation}
U_{TF}^{AL}=\frac{12}{7}\left( \frac{2}{9\pi ^{2}}\right) ^{1/3}\left[
\left( Z_{1}+1\right) ^{7/3}-Z_{1}^{7/3}\right] \left[ S\left( q\right) +%
\frac{q^{2}}{x_{0}\left( q\right) }\right] \frac{e^{2}}{a_{H}}
\end{equation}

\begin{equation}
f_{TF}^{AL}\left( E\right) \simeq \exp \left[ -\frac{\left[ S\left( q\right)
+q^{2}/x_{0}\left( q\right) \right] }{5E^{3/2}}Z_{1}Z_{2}\left[ \left(
Z_{1}+1\right) ^{7/3}-Z_{1}^{7/3}\right] A^{1/2}\right]
\end{equation}
Note that if we set $q=0$ in all the above $SL$ and $AL$ formulas we recover
the respective ones for neutral targets\cite{lioliosluna}. However, the $%
f_{TF}^{SL}\left( E\right) $ SEF given by Eq. $\left( \ref{fslion}\right) $
must be used with caution. It has been derived by truncating Baker's series
but at very low energies and moderately heavy nuclei that approximation
begins to break down. Its actual validity will be investigated along with
the relativistic effects in Sec.III.

We can assess the quality of the constraint imposed by our model on the
possible values of the screening energies and the associated SEFs by
studying Fig. 3 and Fig. 4. In the former, the SL and AL screening energy $%
U_{TF}^{(SL,AL)}$ for the most important astrophysical nuclear reactions of
the $CNO$ bi-cycle are plotted with respect to the degree of ionization of
the TF positive ionic target . Along the whole profile of the $q$ values the
upper $\left( AL\right) $ and lower $\left( SL\right) $ screening energy
limits differ by roughly$\,100\,eV$ (or less) which is a very satisfactory
constraint for most practical purposes.

On the other hand in Fig.4 we can observe the $SL/AL$ $\,SEFs\,\,$for the
most important astrophysical nuclear reaction of the $CNO$ bi-cycle, namely $%
N^{14}\left( p,\gamma \right) O^{15},\,$with respect to the center-of-mass
energy $E$ in the region of the solar Gamow peaks $\left( E_{GP}=27.2\text{ }%
keV\right) $ for various degrees of ionization. Naturally, the more ionized
the atomic target the smaller the screening energy (see also Fig. 3) and,
consequently, the smaller the respective SEF at a particular energy. The
quality of the SL/AL constraint on the SEF is very good as has already been
indicated by the corresponding screening energy limits (see Fig.3.)

Before we proceed to study relativistic effects, we should consider the
influence of thermal and exchange effects. It should be emphasized that we
have defined the sudden limit for a cold electron gas $\left( T=0\right) $
disregarding all exchange effects which, on the contrary, have been taken
into account in the derivation of the total TF binding energy $%
E_{TF}^{tot}\left( q\right) .$ $\,$Of course, the latter also includes the
total potential energy $E_{TF}^{\left( n,e\right) }$ between the nucleus and
the electrons and the total kinetic energy $K_{TF}^{tot}$ of the bound
electrons. Actually, the exchange energy $E_{TF}^{\left( e,e\right) }$is
expected to lower the relative collision energy as recently shown in a
relevant study\cite{lioliosepj}. For a neutral atom the above components of
the total binding energy $E_{TF}^{tot}\left( q\right) $ can be written\cite
{flugge} : 
\begin{equation}
K_{TF}^{tot}=+20.99Z^{7/3},E_{TF}^{\left( e,e\right)
}=+7.00Z^{7/3},E_{TF}^{\left( n,e\right) }=-48.99Z^{7/3}  \label{kee}
\end{equation}
where we observe that exchange effects play a less important screening role
than the kinetic energy ones.

It is now obvious that the adiabatic limit SEF can be written as

\begin{equation}
f_{TF}^{AL}\left( E\right) =f_{K_{TF}^{tot}}^{AL}\left( E\right) \times
f_{E_{TF}^{\left( e,e\right) }}^{AL}\left( E\right) \times f_{E_{TF}^{\left(
n,e\right) }}^{AL}\left( E\right)  \label{feee}
\end{equation}
where the first two factors of the right-hand side decrease rather than
increase the cross section of the astrophysical fusion reaction.

The above factorization of multi-electron SEFs allows some qualitative
investigation of the phenomenon. Obviously, according to the third factor,
by increasing the binding energy of the electrons we can enhance the cross
section of the reaction . That is for the same number of electrons, the
larger the atomic number the larger the cross section at a particular
collision energy. However, by increasing the atomic number the kinetic
energy of electrons is also increased and according to the first factor that
will cause a decrease in the cross section. On the other hand exchange
effects also increase with the atomic number reducing further the cross
section. Note that the dependence of the SEF on the atomic number is more
complex than the simple monotonous increase indicated by Eq. $\left( \ref
{feee}\right) $ and Eq. $\left( \ref{kee}\right) $. The TF model, being a
statistical one, gives a net dependence of the total energy on the atomic
number proportional to $Z^{7/3}.$ Unfortunately this conceals all shell
effects, which are very important, as each shell carries a different
screening capacity. For example, the K-shell electrons are expected to play
a much more important role than those lying in the L-shell and so on. In
fact the sudden limit SEFs which were derived\cite{lioliosepj} for the first
and second excited states in hydrogen-like atoms, showed that a single
electron of the K-shell increases the relative collision energy four times
more than a single electron of the L-shell. As the screening energy enters
the SEF exponentially (Eq. $\left( \ref{fap}\right) $ ) that translates into
a K-shell SEF being equal to the L-shell SEF raised to the fourth power.
Higher shells can be shown to play an even minor role. Notably, the present
TF approach for the sudden limit disregards the fast response of the
electron cloud due to the combined nuclear molecule \cite{lioliosluna}, thus
obtaining an even more conservative value for the SL screening energy.

In order to study the relative importance of shell effects we can consider
the screening energy generated by the collision of a bare Carbon nucleus $%
\left( Z=6\right) \,$with an atom of the same nuclear charge with only two
electrons in its closed K-shell. Disregarding the fast response of the
two-electron cloud, so that we are consistent with the present TF model, the
Hartree-Fock (HF) screening energy is\cite{lioliosluna}

\begin{equation}
U_{HF}^{SL}=\left[ -2Z\left( Z-5/16\right) +\frac{5}{8}\left( Z-5/16\right)
\right] \frac{e^{2}}{a_{H}}
\end{equation}
with a numerical value of $U_{HF}^{SL}=1760\,eV.$

On the other hand the Thomas-Fermi SL model (Eq.$\left( \ref{uslion}\right) $
) for $q=0$ gives $U_{TF}^{SL}=3191\,eV.$ If we also take into account the
exchange effects disregarded by Eq. $\left( \ref{uslion}\right) \,$the TF
screening energy should be smaller. In any case, according to our models,
more than $50\%$ of the six-electron TF screening energy originates from the
two K-shell electrons, which emphasizes the importance of shell effects.

The TF formulas derived in this paper show a direct dependence of
multi-electron SEFs on the ionization state $\left( q\right) $ of the
reacting atoms. The ionization number $q$ is of course a function of
temperature (see for example the Saha equation\cite{claytonbook}) so that at
very high temperatures the atom is completely ionized (e.g. stellar plasma).
Thus, the cold electron gas assumption is no longer valid when the
temperature rises. For thermal energies of a few $eV$ , corresponding to a
few thousand degrees Kelvin, Eq. $\left( \ref{tf0}\right) \,$should be
replaced by the Marshak-Bethe equation \cite{marshak}:

\begin{equation}
\frac{d^{2}\phi _{MB}\left( x,T\right) }{dx^{2}}=\frac{\phi
_{MB}^{3/2}\left( x,T\right) }{x^{1/2}}\left[ 1+\zeta \left( T\right) \frac{%
x^{2}}{\phi _{MB}^{2}\left( x,T\right) }\right]
\end{equation}
where the dimensionless temperature-dependent parameter is $\zeta \left(
T\right) =10^{-11}Z_{1}^{-8/3}T^{2}$ , with $T$ measured in degrees Kelvin.
The solution of that equation can be found by a perturbation method\cite
{feynman} and reads:

\begin{equation}
\phi _{MB}\left( x,T\right) =\phi _{0}\left( x\right) +\zeta \left( T\right)
\phi _{1}\left( x\right) 
\end{equation}
where the unperturbed term $\phi _{0}\left( x\right) $ is the solution of
Eq. $\left( \ref{tf0}\right) $ and the trial function $\phi _{1}\left(
x\right) $ is approximated by the truncated series $\phi _{1}\left( x\right)
\simeq \left( 4/35\right) x^{7/2}-\left( 4/63\right) S\left( 0\right)
x^{9/2}+...$ Consequently, in our study there is now a temperature dependent
screening correction $U_{e}\left( T\right) \,$to be taken into account.
After some algebra that term for small distances can be written

\begin{equation}
U_{e}\left( T\right) =\frac{4}{35}\zeta \left( T\right) \frac{Z_{1}Z_{2}e^{2}%
}{\alpha }\left( \frac{r}{\alpha }\right) ^{5/2}  \label{uet}
\end{equation}
Actually, in obtaining Eq. $\left( \ref{uslion}\right) \,$we disregarded
radially dependent terms arising from Baker's expansion since even the
largest of those neglected terms, given by

\begin{equation}
U_{TF}^{SL}\left( r\right) =\frac{4}{3}\frac{Z_{1}Z_{2}e^{2}}{\alpha }\left( 
\frac{r}{\alpha }\right) ^{1/2}  \label{uer}
\end{equation}
is much smaller than the constant screening shift throughout the tunnelling
region. By comparing Eq. $\left( \ref{uet}\right) \,$and Eq. $\left( \ref
{uer}\right) $ it is obvious that for temperatures of several thousand
degrees Kelvin (so that $kT<<e\Phi \left( r_{0}\right) $ ) the
multi-electron screening enhancement of astrophysical reactions is the same
as in a cold electron gas.

\section{Relativistic effects}

We can now proceed to include relativistic corrections to screening energies
and the respective SEFs, which are expected to become important when
studying the screening effects of lower shell electrons on a heavy atomic
nucleus. For example a K-shell electron of an Oxygen atom has a Bohr
velocity of $u_{e}=0.058c,$ which indicates that relativity may further
modify the screening effect and therefore warrants further investigation.
For simplicity, in the study that follows we will focus on neutral targets,
as any degree of ionization will simply diminish the associated relativistic
corrections.

The sudden limit can be obtained by using the Dirac-Hartree-Fock-Slater
(DHFS) calculations of Salvat et al\cite{salvat}, which provide us with
screening functions accounting for relativistic effects. In that work
Moliere's approximation was used and the corresponding $B_{i}\left( Z\right)
,\beta _{i}\left( Z\right) $ parameters were tabulated for $Z=1-92\,$.
Therefore, when relativistic effects are considered, the SL screening shift
can be written as

\begin{equation}
U_{DHFS}^{SL}=-4\left( \frac{2}{9\pi ^{2}}\right)
^{1/3}Z_{1}^{4/3}Z_{2}\left[ \sum_{i=1}^{3}B_{i}\left( Z_{1}\right) \beta
_{i}\left( Z_{1}\right) \right] \frac{e^{2}}{a_{H}}  \label{usldhfs}
\end{equation}
Unfortunately, Eq. $\left( \ref{usldhfs}\right) $ is an very rough
approximation. Moliere's approach was previously shown to underestimate the
non-relativistic (neutral atom) SL screening energy by $14\%.$ In our study
we have used the relativistic and non-relativistic screened Moliere
potentials given in Ref. \cite{moliere} and Ref. \cite{salvat}, respectively$%
.$ Although they both satisfy the condition that relativistic effects result
in more bound systems none of them is reliable close to the nucleus. For
example Eq. $\left( \ref{usldhfs}\right) $ for the reaction $C^{13}\left(
p,\gamma \right) N^{14}\,$yields $U_{DHFS}^{SL}=406eV\,$which is even lower
than the non-relativistic SL\ screening energy given by Eq. $\left( \ref
{uslion}\right) $. Therefore, the DHFS screening energy turns out to be
inappropriate as a lower bound of the relativistic screening energy.

The most accurate way to calculate the screened Coulomb potential would be
to use the Vallarta-Rosen relativistic TF equation \cite{valarta}

\begin{equation}
\frac{d^{2}\phi _{rel}\left( x\right) }{dx^{2}}=\frac{\phi
_{rel}^{3/2}\left( x\right) }{\sqrt{x}}\left[ 1+\lambda \frac{\phi
_{rel}\left( x\right) }{x}\right] ^{3/2}  \label{vr}
\end{equation}
with $\lambda =3\times 10^{-5}Z_{1}^{4/3}.$ The V-R potential has been
approximated with respect to the distance $x_{c}=\lambda +S\left( 0\right)
\lambda ^{2},$ by a two-branch function\cite{hill} . Namely, for small $x$
and $x>x_{c}$

\begin{equation}
\phi _{rel}^{out}\left( x\right) =\sum_{n=0}^{\infty }\left[ \lambda
^{n}\phi _{n}\left( x\right) \right] -1.206\lambda ^{3/2}\left( 1+0.553\ln
Z\right)
\end{equation}
where $\phi _{0}\left( x\right) $ is the non-relativistic Baker's expansion
given by Eq. $\left( \ref{baker}\right) $ and

\begin{equation}
\phi _{1}\left( x\right) =-6x^{1/2}+11.733x+5S\left( 0\right) x^{3/2}+...
\end{equation}

\begin{equation}
\phi _{2}\left( x\right) =\frac{1}{2}x^{1/2}-\frac{21}{4}S\left( 0\right)
x^{1/2}+...
\end{equation}
The only significant terms added to Baker's expansion through the sum of Eq. 
$\left( 32\right) $ are those raised to powers $n=0,1.$ The remaining terms $%
\left( n>1\right) $ are negligible throughout the tunneling region as they
are multiplied by the very small number $\lambda ^{n}$ and therefore vanish
at distances smaller than the classical turning point (see also previous
section and the Appendix in Ref. \cite{lioliosluna}.

For distances shorter than $x_{c}$ (but larger than the minimum internuclear
distance given by $x_{n}=r_{n}/\alpha )$ the approximation is 
\begin{equation}
\phi _{rel}^{in}\left( x\right) =1+sx+\frac{3}{2}\lambda ^{1/2}\left[ \left(
x_{n}-x\right) +x\ln \left( x/x_{n}\right) \right] +\lambda ^{3/2}\left[
\left( \frac{x-x_{n}}{x_{n}}\right) -\ln \left( \frac{x}{x_{n}}\right)
\right] +...
\end{equation}
where $s$ is given by

\begin{equation}
s=S\left( 0\right) -\lambda ^{1/2}\ln Z-0.1388\lambda ^{1/2}+11.733\lambda
+5.157S\left( 0\right) \lambda ^{3/2}+...
\end{equation}
Instead of seeking an approximate analytic formula for the relativistic $SL$
SEF we have numerically evaluated the quantity given by Eq. $\left( \ref{fex}%
\right) \,$for the non-relativistic and relativistic regime. This approach
will also provide a measure of the accuracy of the non-relativistic $SL$ SEF
analytic formula given by Eq. $\left( \ref{fslion}\right) .$

In Fig. 5 we have plotted the sudden limit $SEFs$ for the most important
astrophysical nuclear reactions of the $CNO$ bi-cycle with respect to the
center-of-mass energy $E$ in the region of the solar Gamow peaks $\left(
E_{GP}\right) $ with and without relativistic corrections (middle and
lowermost curves, respectively), obtained by integrating numerically the
penetration factors appearing in Eq. $\left( \ref{fex}\right) $. In the same
plot we have also included the non-relativistic SEF (uppermost curves)
obtained through the analytic formula given by Eq. $\left( \ref{fslion}%
\right) $. We observe that the analytic formula overestimates the $\left(
CNO\right) $ $SEFs$ by roughly $2\%$ at the Gamow peak energies, with the
discrepancy becoming larger at lower energies. In nuclear astrophysics
experiments, as far as astrophysical factors are concerned, energies lower
than the Gamow peaks are not needed for most practical purposes, unless of
course there are suspicions about the existence of resonances. If resonances
can be ruled out, or the experimental energies are higher than the Gamow
peaks (as expected to happen in future experiments) then the analytic $%
\left( SL\right) $ $SEF$ formula of Eq. $\left( \ref{fslion}\right) \,$is a
very reliable lower bound of our non-relativistic constraint.

In the same figure, regarding the lowermost and middle curves of each
reaction, although it is obvious that the SL relativistic curves are almost
indistinguishable from the non-relativistic ones , their divergence is
increased when moderately heavy atoms are considered at very small energies.
Admittedly, such small effects can be neglected at the currently attained
experimental energies (or at the even lower energies expected to be attained
in the future). Consequently, in any case, the non-relativistic $SL$ SEF is
lower than the relativistic one and therefore can safely play the role of
the lower bound of our constraint.

The upper (adiabatic) limit (AL) is expected to be more reliable as it has
been calculated under well established models. In fact the total (neutral
atom) TF energy can be given by:

\begin{equation}
E_{TF}^{rel}\left( Z\right) =Z^{7/3}\sum_{n=0}^{\infty }\left(
c_{n}+k_{n}\ln Z\right) Z^{2n/3}
\end{equation}
where the coefficients $c_{n},k_{n}$ have been tabulated in Ref. \cite{hill}$%
.$ Following the treatment used in Ref. \cite{assen}  the relativistic AL
screening energy for proton induced reactions will be:

\begin{equation}
U_{RTF}^{AL}=\left( Z_{1}+1\right) ^{7/3}\sum_{n=0}^{\infty }\left[
c_{n}+k_{n}\ln \left( Z_{1}+1\right) \right] \left( Z_{1}+1\right)
^{2n/3}-Z_{1}^{7/3}\sum_{n=0}^{\infty }\left[ c_{n}+k_{n}\ln \left(
Z_{1}\right) \right] Z_{1}^{2n/3}  \label{ualrel}
\end{equation}

The corresponding relativistic $AL$ $SEF$ can be found using Eq. $\left( \ref
{fap}\right) $ and the above screening energy. In Fig. 6 we have plotted the
adiabatic limit $SEFs$ for the most important astrophysical nuclear
reactions of the $CNO$ bi-cycle with respect to the center-of-mass energy $E$
in the region of the solar Gamow peaks $\left( E_{GP}\right) $ with and
without relativistic corrections (neutral targets). Upper and lower curves
always indicate relativistic and non-relativistic regimes, respectively. As
expected the divergence between the relativistic and the non-relativistic
curves is an increasing function of the atomic number and a decreasing
function of the relative collision energy. The screening energy given by Eq. 
$\left( \ref{ualrel}\right) $ and the SEFs originating from it (through Eq. $%
\left( \ref{fap}\right) )$ are the highest possible values and constitute
the upper bounds of the constraints derived in this paper

\section{Conclusions}

The astrophysical factors $S\left( E\right) $ for nuclear reactions ignited
during advanced stages of stellar evolution (solar center, upper main
sequence, classical novae etc.) have been obtained by performing
measurements in the laboratory well above the Gamow peak, and extrapolating
to lower energies. Any attempt to increase the accuracy of $S\left( E\right) 
$ by lowering the experimental energies introduces a significant error as
multi-electron screening effects enhance the cross sections at astrophysical
energies, thus yielding an overestimated zero-energy astrophysical factor $%
S\left( 0\right) .$ The tool for correcting the respective data is the SEF
which, in the present work, is derived both analytically and numerically for
the most important nuclear reactions of the CNO bi-cycle by means of
Thomas-Fermi model. By means of that model lower $\left( SL\right) $ and
upper $\left( AL\right) $ limits are derived for the possible values of the
screening energies and the respective SEFs imposing a very reliable
constraint on the corrections which are necessary before fitting the
bare-nucleus astrophysical factor polynomial to the experimental data.

Moreover, we have also investigated the effects of the degree of ionization $%
\left( q\right) $ of the reacting atoms deriving analytically the functional
dependence of SEF on it. As expected, the SEF is a decreasing function of $q$
, which is again well constrained by lower and upper limits. Other effects
studied here, are exchange, relativistic and thermal ones. The actual
contribution of exchange effects is analysed while relativistic and thermal
effects are shown to be negligible for most practical purposes. Finally the
importance of shell effects, an aspect the TF model cannot take into
account, is underlined by showing that the lower shell are mostly
responsible for the screening effect while higher ones play a minor role
even in multi-electron atomic systems.

{\bf ACKNOWLEDGMENTS}

This work was financially supported by the Greek State Grants Foundation
(IKY) under contract \#135/2000.

\bigskip FIGURE CAPTIONS

Figure 1. The initial slope $S\left( q\right) =\phi ^{^{\prime }}\left(
0\right) $ of the TF function $\phi \left( x\right) $ plotted with respect
to the degree of ionization $q=1-N/Z$ of the TF positive ion.

Figure 2. The ionic radius $x_{0}\left( q\right) \,$\thinspace of the TF
positive ion plotted with respect to the degree of ionization $q=1-N/Z$ .

Figure 3. The SL and AL screening energy $U_{TF}^{(SL,AL)}$ for the most
important astrophysical nuclear reactions of the $CNO$ bi-cycle plotted with
respect to the degree of ionization $q=1-N/Z$ of the TF positive ion. The
upper (lower) solid curve represents the enhancement of the $C^{13}\left(
p,\gamma \right) N^{14}$ reaction as calculated by the TF sudden (adiabatic)
limit. Likewise, the dashed curves stand for the $N^{14}\left( p,\gamma
\right) O^{15}$ reaction, while the dotted ones for the $O^{16}\left(
p,\gamma \right) F^{17}$ reaction$.$ Upper and lower curves always indicate
sudden and adiabatic limits respectively.

Figure 4.

The $SEFs\,\,$for the most important astrophysical nuclear reactions of the $%
CNO$ bi-cycle $N^{14}\left( p,\gamma \right) O^{15}$ reaction$\,$with
respect to the center-of-mass energy $E$ in the region of the solar Gamow
peaks $\left( E_{GP}=27.2\text{ }keV\right) $ for various degrees of
ionization $q=1-N/Z$. The lower (upper) solid curve represents the
enhancement of the reaction as calculated by the TF sudden (adiabatic) limit
for neutral atoms (Eqs. $\left( \ref{fslion}\right) $ and \ref{falion}).
Likewise, the dashed curves stand for a degree of ionization $q=0.3$, while
the dotted ones for $q=0.8.$ Lower and upper curves always indicate sudden
and adiabatic limits, respectively.

Figure 5.

The sudden-limit $SEFs$ for the most important astrophysical nuclear
reactions of the $CNO$ bi-cycle with respect to the center-of-mass energy $E$
in the region of the solar Gamow peaks $\left( E_{GP}\right) $ with and
without relativistic corrections (neutral targets). The solid curves
represent the enhancement of the $C^{13}\left( p,\gamma \right) N^{14}$
reaction $\left( E_{GP}=24.5\text{ }keV\right) $ with the lowermost and
middle curves standing for the non-relativistic and relativistic SEFs,
respectively, obtained by integrating numerically the penetration factors
appearing in Eq. $\left( \ref{fex}\right) $. The uppermost solid curve
represents the non-relativistic SEF obtained through the analytic formula
given by Eq. $\left( \ref{fslion}\right) $. Likewise, the dashed curves
stand for the $N^{14}\left( p,\gamma \right) O^{15}$ reaction $\left(
E_{GP}=27.2\text{ }keV\right) $, while the dotted ones for the $O^{16}\left(
p,\gamma \right) F^{17}$ reaction $\left( E_{GP}=29.8\text{ }keV\right) ,$
with their lowermost, middle and uppermost curves having the same meaning as
in the $C^{13}\left( p,\gamma \right) N^{14}$ reaction.

Figure 6.

The adiabatic limit $SEFs$ for the most important astrophysical nuclear
reactions of the $CNO$ bi-cycle with respect to the center-of-mass energy $E$
in the region of the solar Gamow peaks $\left( E_{GP}\right) $ with and
without relativistic corrections (neutral targets). The upper (lower) solid
curve represents the relativistic (non-relativistic) enhancement of the $%
C^{13}\left( p,\gamma \right) N^{14}$ reaction as calculated by the TF
model. Likewise, the dashed curves stand for the $N^{14}\left( p,\gamma
\right) O^{15}$ reaction, while the dotted ones for the $O^{16}\left(
p,\gamma \right) F^{17}$ reaction$.$ Upper and lower curves always indicate
relativistic and non-relativistic regimes, respectively.

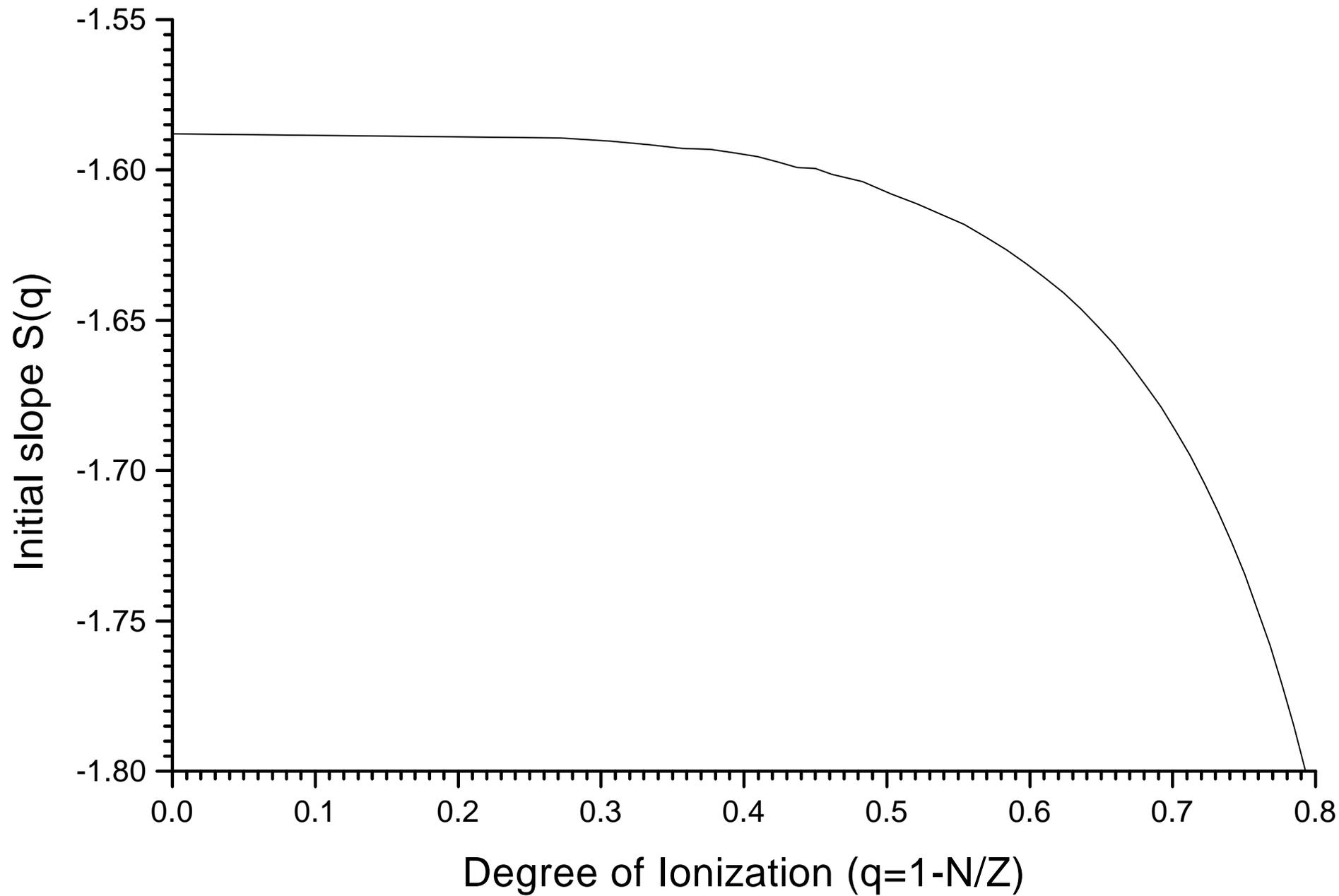
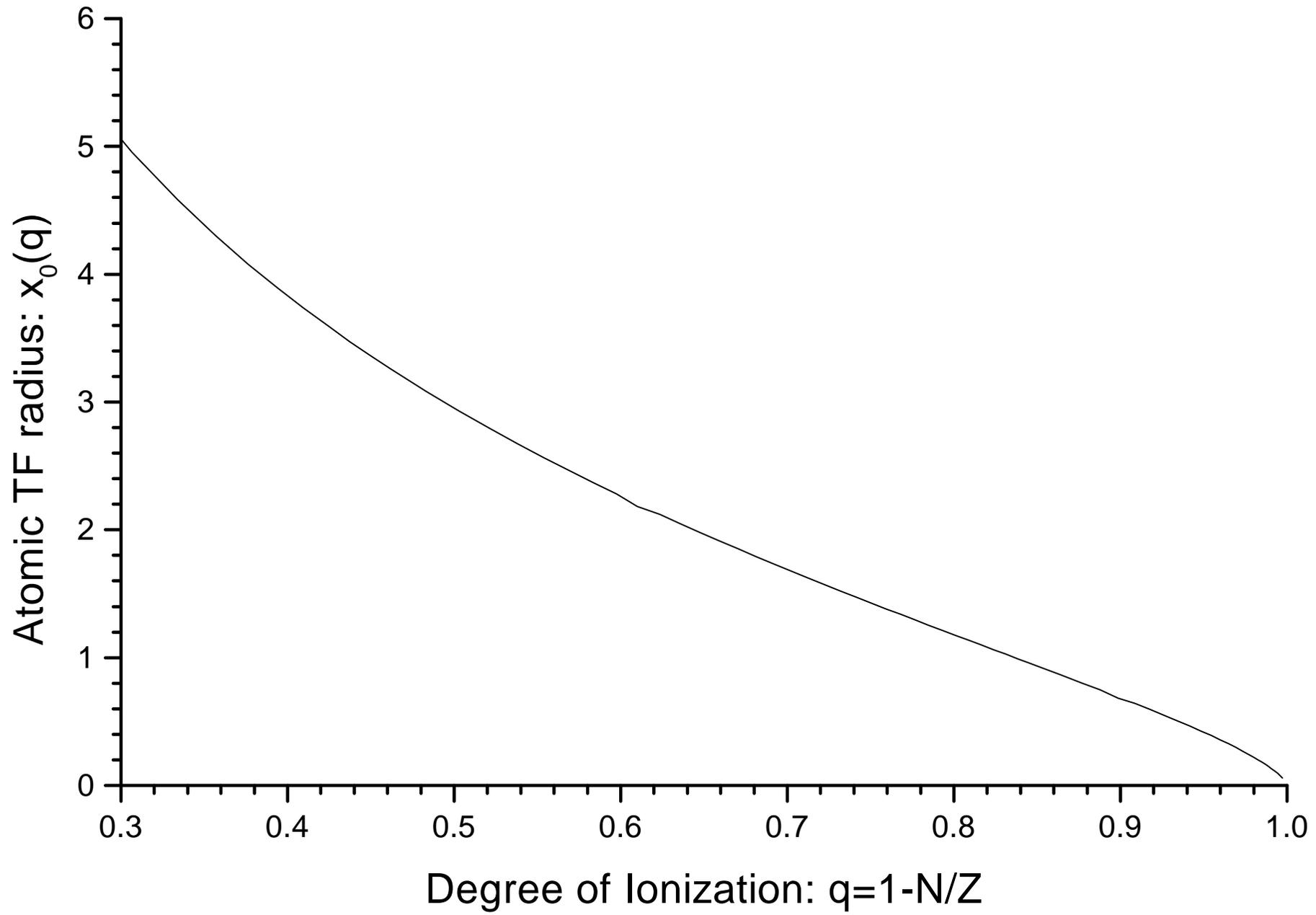
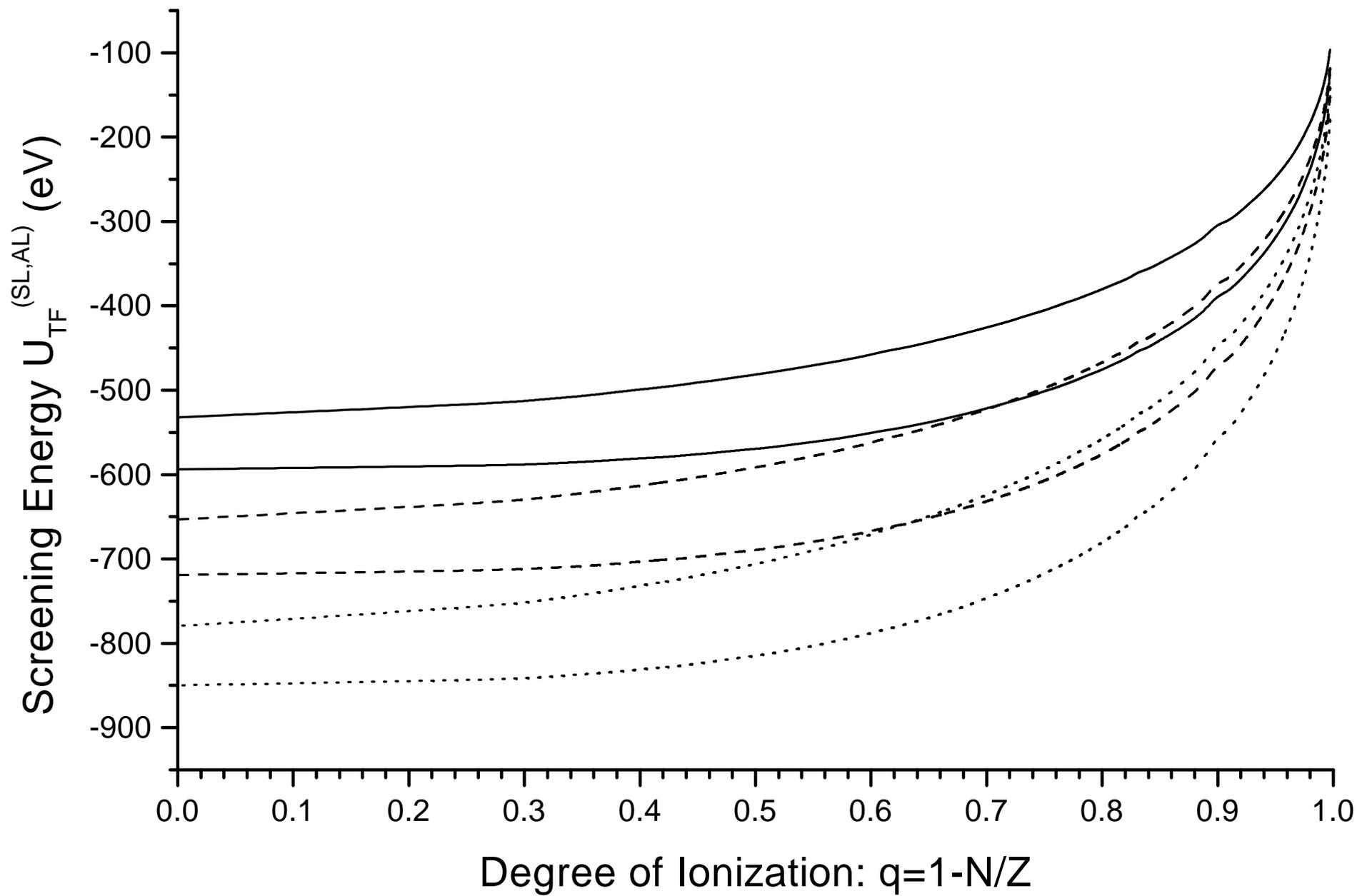
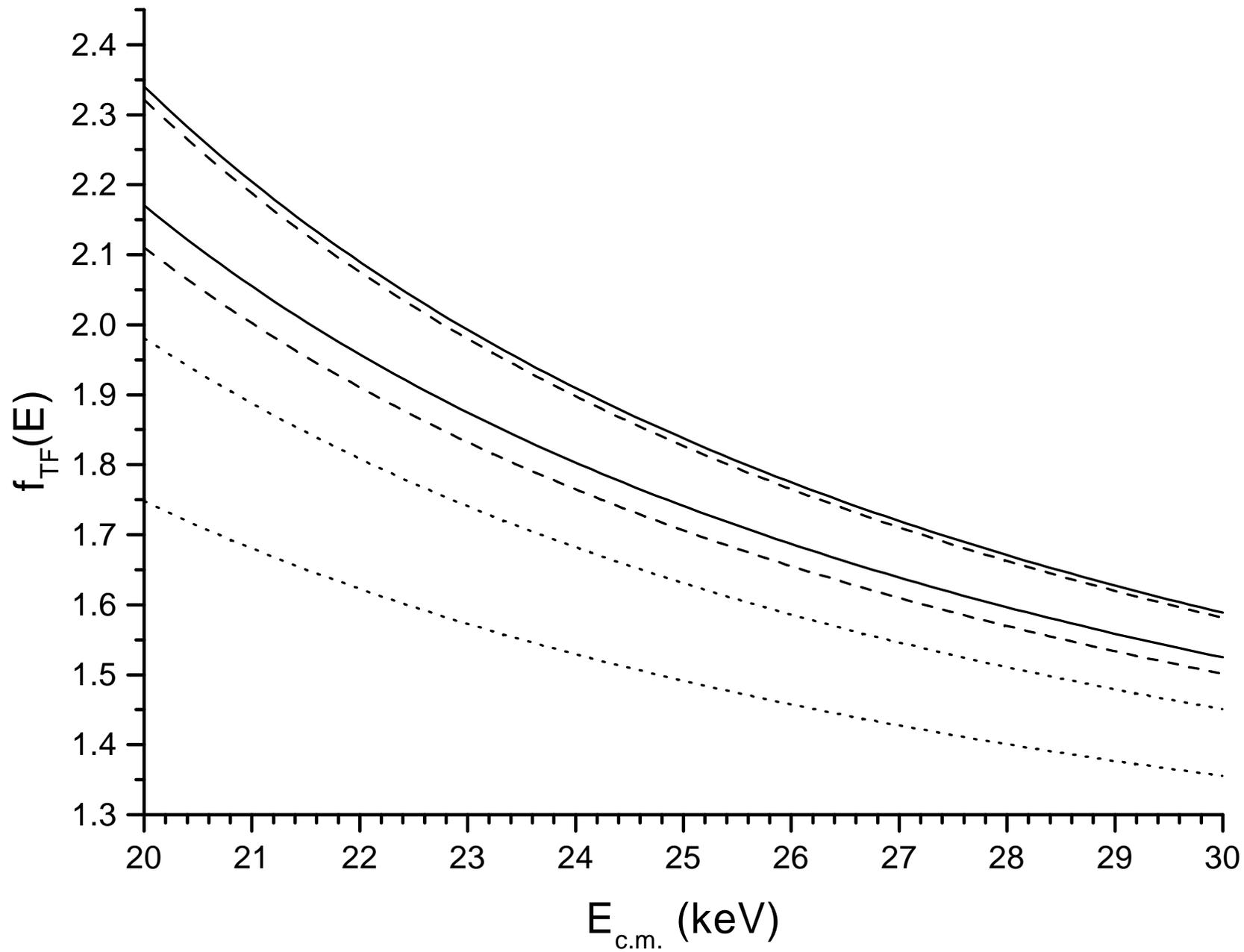
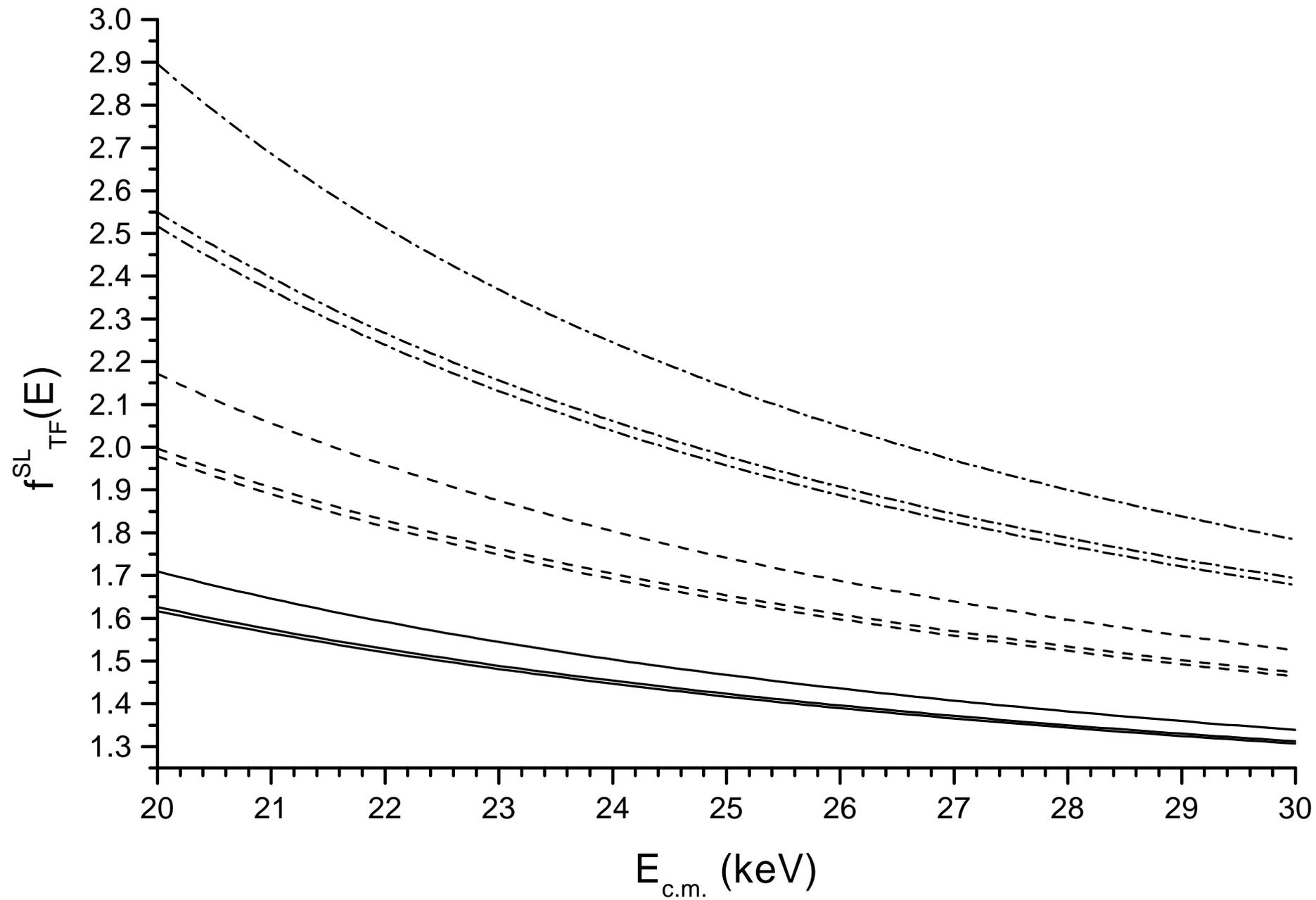
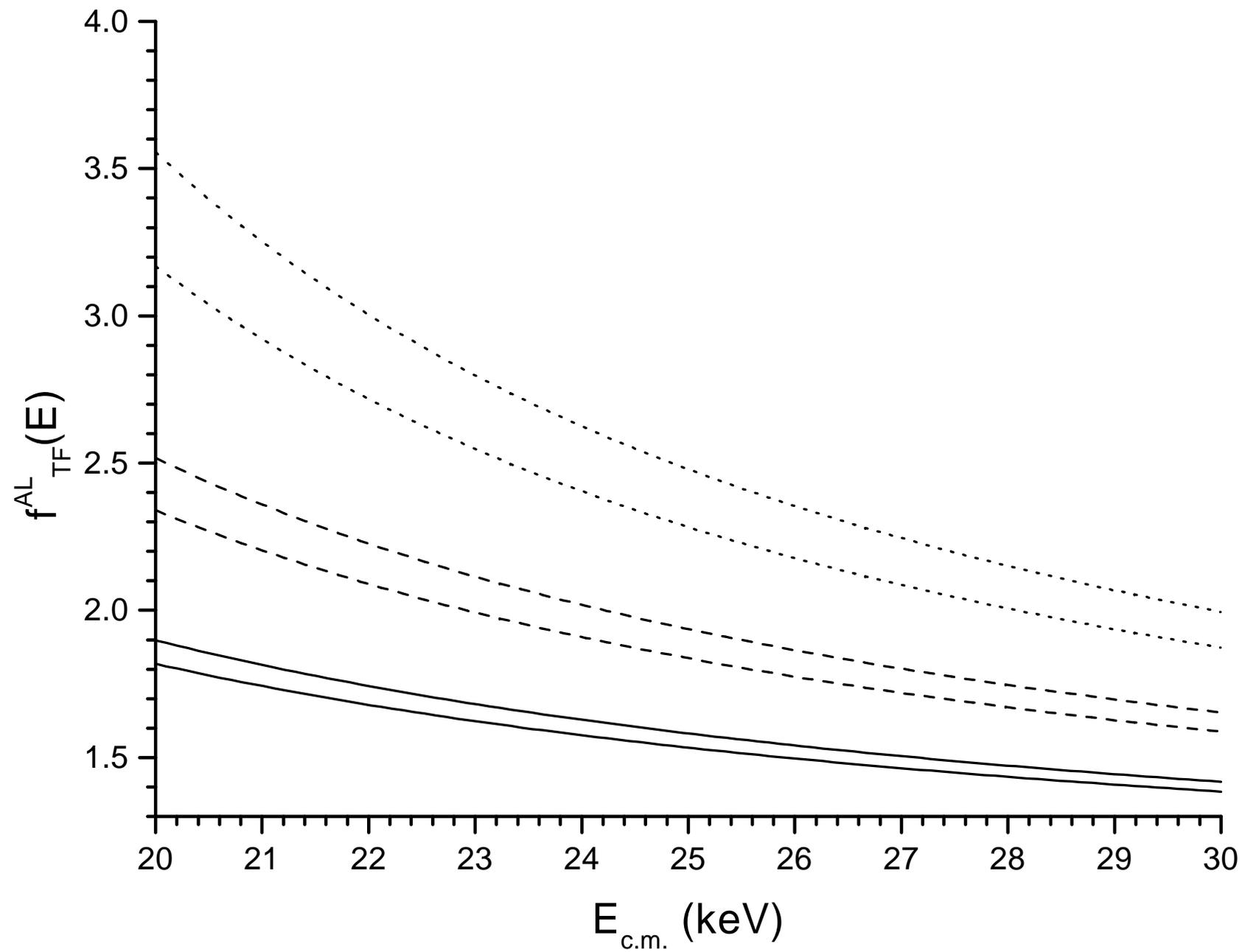
\end{document}